\documentstyle[12pt]{article}
\begin{document}
\title{Low-energy scattering in four nucleon systems. Method of
Cluster Reduction} 
\author{S.L. Yakovlev\thanks{{\it E-mail address:}  
yakovlev@mph.phys.spbu.ru}, I.N. Filikhin}
\date{ }
\maketitle
\begin{center}
{\it Department of Mathematical and Computational
Physics, St. Petersburg State University,  
198904 St. Petersburg, Petrodvoretz, Ulyanovskaya Str.
1, Russia}  
\end{center}
\section{Cluster reduction of YDE}
The elastic and rearrangement processes in the four-particle system with
two clusters in the 
initial and final states can be treated adequately on the base 
of Yakubovsky differential
equations (YDE) \cite{MYa1}, \cite{MYa2}, \cite{MYaG}
$$
(H_{0}+V_{a_3}-E)\Psi_{a_{3}a_{2}}+V_{a_3}\sum_{({c_3}\neq {a_3})\subset 
{a_2}}\Psi_{c_{3}a_{2}} =  
-V_{a_3}\sum_{{d_2}\neq {a_2}}\sum_{({d_3}\neq {a_3})\subset 
{a_2}}\Psi_{d_{3}d_{2}}.  
$$
For the two cluster collisions the YDE admit a further
reduction. Let $H_0 = T_{a_2} + T^{a_2}$ 
be the separation of the kinetic energy operator in the intrinsic
$T_{a_2}$ with respect to the clusters of $a_2$ part and the kinetic
energy $T^{a_2}$ of the relative motion of $a_2$ clusters. The
cluster reduction procedure consists in expansion of the components
$\Psi_{a_{3}a_{2}}$ along the basis of the solutions to the Faddeev
equations (FE) for subsystems of partition $a_2$
$$
(T_{a_2} + V_{a_{3}})\psi_{{a_2},k}^{a_3} + V_{a_3}
\sum_{({c_3}\neq {a_3})\subset {a_2}}\psi_{{a_2},k}^{c_3} = 
\varepsilon_{a_2}^{k}\psi_{{a_2},k}^{a_3}.
$$ 
The expansion has the form 
\begin{equation}
\Psi_{{a_3}{a_2}}({\bf X}) = \sum_{k=0}^{\infty}
\psi_{{a_2},k}^{a_3}({\bf x}_{a_2})
F_{a_2}^{k}({\bf z}_{a_2}).
\label{expan}
\end{equation}
Here, the unknown amplitudes $F_{a_2}^{k}({\bf z}_{a_2})$ depend
only on the relative position vector ${\bf z}_{a_2}$ between the clusters
of the 
partition $a_2$ and by ${\bf x}_{a_2}$ are denoted intrinsic with respect
to clusters of partition $a_2$ coordinates. The basis of the solutions of
FE is complete but not 
the orthogonal one \cite{Yak1}, \cite{Yak2} due to not Hermitness of FE.
The biorthogonal basis is formed by the solutions of conjugated to FE
equations 
$$
(T^{a_2} + V_{a_{3}})\phi_{{a_2},k}^{a_3} + 
\sum_{({c_3}\neq {a_3})\subset {a_2}}V_{c_3}\phi_{{a_2},k}^{c_3} = 
\varepsilon_{a_2}^{k}\phi_{{a_2},k}^{a_3}.
$$  
Introducing expansion for $\Psi_{{a_3}{a_2}}({\bf X})$ into YDE and
projecting 
 onto the elements of biorthogonal basis 
$\{\phi_{{a_2},k}^{a_3}({\bf x}_{a_2})\}$
lead to resulting reduced YDE (RYDE)\cite{YakFil1}, \cite{YakFil2} for
$F_{a_2}^{k}({\bf z}_{a_2})$ 
\begin{equation}
(T^{a_2} - E + \varepsilon_{a_2}^{k})F_{a_2}^{k} =   
-\sum_{{a_3}\subset {a_2}}\langle 
\phi_{{a_2},k}^{a_3}|V_{a_3}\sum_{{d_2}\neq {a_2}}\sum_{({d_3}\neq
{a_3})\subset {a_2}}\sum_{l\geq
0}\psi_{{d_2},l}^{{d_3}}F_{d_2}^{l}\rangle ,
\label{RYDE}
\end{equation}
where the brackets $\langle .|. \rangle $
mean the integration over ${\bf x}_{a_2}$.
The boundary conditions for $F_{a_2}^{k}({\bf z}_{a_2})$ have the 
following {\em two body} form as $|{\bf z}_{a_2}| \rightarrow \infty $
$$
F_{a_2}^{k}({\bf z}_{a_2}) \sim \delta_{k0}[\delta_{{a_2}{b_2}}
\exp i({\bf p}_{a_2},{\bf z}_{a_2}) + 
{\cal A}_{{a_2}{b_2}}\frac{\exp{i\sqrt{E-\varepsilon_{a_2}^{0}}
|{\bf z}_{a_2}|}}{|{\bf z}_{a_2}|}], 
$$
where the index $b_2$ corresponds to the initial state, and ${\bf p}_{a_2}$
is the conjugated to ${\bf z}_{a_2}$ momentum. The charged particles case
can be treated in framework of YDE formalism by adjusting the Coulomb
potentials to the kinetic energy operator $H_{0}$ and by replacing the
plane and spherical waves in the asymptotics by respective Coulomb
modifications \cite{MYaG}.  
\section{Application to low-energy scattering in four nucleon system}
RYDE (\ref{RYDE}) after suitable partial wave decomposition 
become one dimensional in variable $|{\bf z}_{a_2}|$. We solve 
numerically these equations by means of finite-difference approximation in
$|{\bf z}_{a_2}|$ variable 
, spline expansion of integrand in right hand side and truncation of
summation over $l$ by finite number $N$. In all the cases, the satisfactory
convergence was observed with parameter $N$ not exceeding 20 what supports
the efficiency of the expansion (\ref{expan}). The maximal size of linear
system to solve was of the order $10^{5}$, so that the calculations were
performed on a standard workstation. 
We have used MT I-III model
with parameters from \cite{Friar} for $NN$ forces.

First group of results we are presenting concerns the isospin
approximation ({\it i. e.} neglecting Coulomb interaction). 
The values of
channel scattering lengths for nucleon scattered off three-nucleon
cluster presented in Table 1 are in agreement with results of Grenoble group
obtained by a direct discretization of YDE. Note, that the $T=1$
channels correspond to singlet and triplet $n-{^3}$H scattering.

Second group of results is more realistic in view of taking account of
Coulomb interaction between protons. In Table 2 we collect our results for
$p-{^3}$H ($^{2S+1}$A$_{pt}$) and $p-{^3}$He ($^{2S+1}$A$_{ph}$)
elastic scattering lengths with data obtained on the base of
Resonating Group Method (RGM) \cite{Vasil} calculations end experimental
values from \cite{Exp}. Last two rows of Table 2 show the position $E_{r}$
and weight $\Gamma $ in MeV of $^4$He nucleus $0^{+}$ resonance (measured
relatively to $n-{^3}$He threshold) extracted from low energy behavior of
calculated  $p-{^3}$H phase-shift.
The last Table 3 contain results of calculations of scattering lengths
for $n-{^3}$He ($^{2S+1}$A$_{nh}$) and $^2$H$-^{2}$H ($^{2S+1}$A$_{dd}$)
scattering. Due to open rearrangement channels the scattering lengths in
these cases have nontrivial imaginary part. 
\section*{Acknowledgement}
This work was partially supported by Russian Foundation for Basic Research
grant No. 98-02-18190. Authors would like to thank G.M. Halle for
providing his preliminary result of R-matrix analysis of experimental data
for pentaplet $^2$H$-{^2}$H scattering.
\newpage 
\section*{Tables}
\noindent 
\begin{table}[ht]
\caption{$S-T$ channel $N-NNN$ scattering lengths (in fm)}
\begin{center}
\begin{tabular}{|c|c|c|c|}
\hline 
$S$   &  $T$   & \cite{Carbon}, \cite{Ciesiel-Fonseca}& our \\
\hline\hline  
0   &   0   &    14.75      & 14.7    \\
1   &   0   &    3.25       & 3.2      \\
0  & 1      &     4.13      & 4.0      \\ 
1  &   1    &     3.73      & 3.6      \\
\hline 
\end{tabular}
\end{center}
\end{table}
\begin{table}[ht] 
\caption{Singlet and triplet $p-{^3}$H and $p-{^3}$He scattering lengths
(in fm)}
\begin{center}
\begin{tabular}{|c|c|c|c|c|c|c|}
\hline
     &$^1$A$_{pt}$ &$^3$A$_{pt}$ & $^1$A$_{ph}$ & $^3$A$_{ph}$ 
                                                &$E_{r}$ &$\Gamma$ \\
\hline \hline 
our          & -22.6         & 4.6         & 8.2    & 7.7  &0.15&0.3\\
\cite{Vasil} & -21.46        &             &        &      &0.12&0.26\\
\cite{Exp} & & & 10.8$\pm $ 2.6 & 8.1$\pm $0.5 & & \\
\cite{Exp0+} &               &        &    & &0.3$\pm $0.05 & 0.27$\pm $0.05\\
\hline
\end{tabular}
\end{center}
\end{table}
\begin{table}[ht] 
\caption{Singlet and triplet $p-{^3}$H and singlet and pentaplet
$^2$H$-{^2}$H scattering lengths (in fm)}
\begin{center}
\begin{tabular}{|c|c|c|c|c|}
\hline
     &$^1$A$_{nt}$ &$^3$A$_{nt}$ & $^1$A$_{dd}$ & $^5$A$_{dd}$\\
\hline \hline 
our          & 7.5-4.2i      & 3.0+0.0i    & 10.2-0.2i    & 7.5  \\
\cite{Vasil} & 7.25-3.92i    &             &              &      \\
\cite{Halle} &               &             &         & 6.68-0.135i\\
\hline
\end{tabular}
\end{center}
\end{table}

\end{document}